\documentclass[manuscript]{aastex}

\def\VEC#1{\mbox{\boldmath $#1$}}

\slugcomment{Not to appear in Nonlearned J., 45.}

\shorttitle{Causal electromagnetic extraction of black hole energy}
\shortauthors{Koide et al.}

\begin{document}


\title{ Causal extraction of black hole rotational energy by various 
kinds of electromagnetic fields} 


\author{Shinji Koide and Tamon Baba}
\affil{Department of Physics, Kumamoto University, 
2-39-1, Kurokami, Kumamoto, 860-8555, JAPAN}




\begin{abstract}
Recent general relativistic MHD simulations have suggested that relativistic jets
from active galactic nuclei (AGNs) have been powered by rotational energy of
central black holes. Some mechanisms of extraction of black hole rotational
energy have been proposed, like the Penrose process, Blandford-Znajek mechanism,
MHD Penrose process, and superradiance. The Blandford-Znajek mechanism is the 
most promising 
mechanism for the engines of the relativistic jets from AGNs. However, an intuitive
interpretation of this mechanism with causality is not clarified yet, while
the Penrose process has a clear interpretation for the causal energy extraction
from the black hole with negative energy. In this paper, we present a formula to
build physical intuition so that
in the Blandford-Znajek mechanism as well as other electromagnetic processes, 
negative electromagnetic energy
plays an important role to extract the rotational energy of the black holes causally.
\end{abstract}





\section{Introduction} \label{sec1}
A number of observations suggested that phenomena in most active
regions in the universe are related to black holes.
Some of most active objects in the universe, for example, active galactic
nuclei (AGNs), micro-quasars (black hole binaries), and gamma-ray
bursts (GRBs), emit relativistic jets \citep{biretta99,pearson87,mirabel94,tingay95,kulkarni99}.
It is believed that these relativistic jets are caused by the drastic phenomena 
around the black holes at the centers of these objects.
The possible energy sources of the drastic phenomena are
gravitational energy of the matter falling toward the black hole
and rotational energy of the black hole itself.
Recently, numerical simulations of general relativistic MHD (GRMHD) have
suggested that the relativistic jet is launched from the vicinity
of the black hole, i.e. inside of the ergosphere \citep{koide04,koide06}, and
some long-term simulations showed that the energy
is seemed to be supplied from the rotational energy of the black hole
\citep{mckinney06,mckinney12}.
In these GRMHD simulations,
the black hole rotational energy seems to be extracted through 
the magnetic field flux tubes due to
the so called Blandford-Znajek mechanism \citep{blandford77}.
It was proposed as the mechanism in the force-free condition, 
by which the rotational energy of the
black hole is extracted directly through the horizon along
the magnetic flux tubes. However, in principle, causality
prohibits the outward transportation of
any material, energy, and information across the horizon. 
Thus, as pointed out by \citet{punsly89,punsly90a,punsly90b},
the Blandford-Znajek mechanism seems to be contradictory to the causality.
On the other hand, in the Penrose process, the black hole spin energy is extracted  
causally due to the negative energy-at-infinity (or just called ``energy") of 
a particle caused by fission \citep{penrose69}.
\citet{takahashi90} and \citet{hirotani92} found
the axisymmetric steady-state solution of ideal MHD plasma inflow 
with the negative energy toward the rotating black hole.
When the negative energy of the inflowing plasma in the ergosphere
is swallowed by the black hole, the black hole rotation energy decreases,
that is, the black hole energy is extracted, just like the Penrose process.
The difference between the ideal MHD mechanism and the Penrose process 
is that the negative energy is produced by the magnetic
tension force in the ideal MHD inflow, while in the Penrose process it
is caused by the fission of a particle. This MHD energy extraction mechanism
is called ``MHD Penrose process" (see Table \ref{clasmech}). 
The MHD Penrose process was mimicked and confirmed
by the GRMHD simulations of initially uniform, very strongly
magnetized plasma around a rapidly rotating black hole, which showed that 
the negative energy of plasma is produced quickly in the ergosphere
\citep{koide02,koide03}. However, because of the short time duration of the simulation,
the numerical solution is far from a stationary state.
\citet{komissarov05} performed a long term GRMHD 
simulations with the similar
initial situation of Koide (2003), and confirmed the MHD Penrose process 
in the early stage. Furthermore, he found that the MHD Penrose process is
a transient phenomenon and alternately the outward electromagnetic energy flux
through the horizon stationarily appears almost everywhere with the exception
of a very thin equatorial belt. He remarked that the pure electromagnetic 
mechanism with ideal MHD condition continues to operate to extract 
the rotational energy of the black hole.
Strictly speaking, this electromagnetic mechanism should be distinguished
from the original Blandford-Znajek mechanism because the original mechanism
is derived with the force-free condition, while the electromagnetic
energy extraction mechanism was shown with the ideal MHD simulations.
In this paper, we call the mechanism shown by the simulations ``MHD
Blandford-Znajek mechanism" while the original mechanism is called
``force-free Blandford-Znajek mechanism".
Considering the numerical results,
\citet{komissarov09} discussed the electromagnetic extraction mechanism 
of the black hole energy, including the force-free Blandford-Znajek mechanism, 
MHD Penrose mechanism, and superradiance in the wide view.
However, unfortunately, the convincing explanation with respect to the causality
of these mechanisms, which should also yield the conditions of the mechanisms, 
is not given except for the MHD Penrose process \citep{komissarov09}.
\citet{koide03} pointed out that the force-free Blandford-Znajek mechanism uses the
{\it negative electromagnetic energy-at-infinity} to extract
the spin energy of the black hole.
This point of view was discussed extensively by \citet{krolik05} and \citet{lasota14}
for ideal MHD and force-free Blandford-Znajek mechanisms, respectively.
However, it is often difficult to build physical intuition on the 
MHD/force-free Blandford-Znajek mechanisms with causality. 
Here, we present an intuitive formula for the electromagnetic mechanism of
the energy extraction from the rotating black hole to aid in building the
physical intuition on the mechanisms.
The formula is also applicable to other electromagnetic mechanisms like the 
MHD Penrose process \citep{takahashi90,hirotani92,koide02,koide03}
and super-radiance \citep{press72,teukolsky74,lightman75}.

In section \ref{sec2}, we review the energy and angular momentum transport
of electromagnetic field around the black holes briefly but sufficiently.
In section \ref{sec3},
we explain the electromagnetic mechanisms of the black hole energy extraction,
that is, the force-free Blandford-Znajek mechanism, MHD Blandford-Znajek mechanism, 
and superradiance within causality. 
We summarize our explanation about the energy extraction mechanisms 
from the black hole including the both Blandford-Znajek mechanisms 
in section \ref{sec4}.

\section{Electromagnetic energy and angular momentum transport 
near rotating black hole \label{sec2}}

We review the electromagnetic energy and angular momentum transport
in the space-time $(x^0, x^1, x^2, x^3)$
around a spinning black hole based on so called ``3+1 formalism". The scale
of a small element in the space-time around the rotating black hole is given by
\begin{equation}
ds^2 = g_{\mu \nu} dx^\mu dx^\nu 
=-h_0^2 dt^2 +\sum _{i=1}^3 \left [h_i^2(dx^i)^2 - 2h_i^2 \omega _i dt dx ^i 
\right]   .
\end{equation}
Here, we have $g_{ij}=0 (i \neq j)$, $g_{00}=-h_0^2$, $g_{ii}=h_i^2$, 
$g_{i0}=g_{0i}=- h_i^2 \omega_i$, where Greek indices $(\mu, \nu)$ run from 0 to 3
and Roman indices $(i,j)$ run from 1 to 3. 
Through this paper, we use the natural unit system, where the light speed, electric permittivity,
and magnetic permeability in vacuum are unity: $c=1$, $\epsilon_0=1$, and $\mu_0=1$.
When we define the lapse function $\alpha$ and shift vector $\beta ^i$ by
\begin{equation}
\alpha = \root \of {h_0^2+\sum _{i=1}^3 
\left ( h_i \omega _i \right ) ^2} ,
\verb!    !
\beta ^{i} = \frac{h_i \omega _i}{\alpha }   ,
\label{diff_alpbet}
\end{equation}
the line element $ds$ is written as
\begin{equation}
ds^2=-\alpha ^2 dt^2+\sum _{i=1}^3 (h_i dx^i - \alpha \beta^i dt)^2 .
\label{eqlinelement}
\end{equation}
The determinant of the matrix with elements $g_{\mu \nu}$ 
is given by
$\root \of {- \| g \|} = \alpha h_1 h_2 h_3$, and
the contravariant metric is written explicitly as
\( \displaystyle
g^{00}=- \frac{1}{\alpha ^2}  ,
g^{i0}=g^{0i}=- \frac{\beta^i}{\alpha h_i},
\) and \( \displaystyle
g^{ij} = \frac{1}{h_i h_j} ( \delta ^{ij}
-\beta ^i \beta ^j ),
\)
where $\delta ^{ij}$ is the Kronecker $\delta$ symbol.

The relativistic Maxwell equations are
\begin{eqnarray}
{\nabla _\mu} ^\ast F^{\mu \nu} = 0   ,
\label{eqfa}
\\
\nabla _\mu F^{\mu \nu} 
=- J^\nu   ,
\label{eqam}
\end{eqnarray}
where $\nabla _\nu$ is the covariant derivative, $F_{\mu \nu}$ is the electromagnetic 
field-strength tensor, and $^\ast F^{\mu \nu}$ is the dual tensor of $F_{\mu \nu}$,
$\displaystyle ^\ast F^{\mu \nu} 
\equiv \frac{1}{2} \epsilon^{\mu \nu \lambda \sigma} F_{\lambda \sigma}$
($\epsilon^{\mu \nu \lambda \sigma}$ is the Levi-Civita anti-symmetric tensor,
which is a tensor density of weight -1), and
$J^\nu = (\rho _{\rm e}, J^1, J^2, J^3)$ is the electric 4-current density 
($\rho _{\rm e}$ is the electric charge density)\citep{jackson79}. 
The electric field $E_i$ and the magnetic field $B^i$
are given by $E_i = F_{i0}$ $(i=1,2,3)$ and $B^1=F_{23}$,
$B^2=F_{31}$, $B^3=F_{12}$ or $\displaystyle B^i = \frac{1}{2} \epsilon^{0ijk} F_{jk}
= {^\ast} F^{0i}$, 
respectively.
Using the 4-vector potential 
$A_\mu$, we have $F_{\mu \nu} = \nabla_\mu A_\nu - \nabla_\nu A_\mu 
= \partial_\mu A_\nu - \partial_\nu A_\mu$ because of the symmetry of
Christoffel symbols, $\Gamma_{\mu \nu}^\lambda = \Gamma_{\nu \mu}^\lambda$.

The electromagnetic energy-momentum tensor $T_{\rm EM}^{\mu \nu}$
is given by
\begin{equation}
T_{\rm EM}^{\mu \nu} = {F^\mu}_\sigma F^{\nu \sigma} - 
\frac{1}{4}g^{\mu \nu} F^{\lambda \kappa} F_{\lambda \kappa}   .
\end{equation}
The total energy-momentum tensor $T^{\mu \nu}$ is 
\begin{equation}
T^{\mu \nu} = p g^{\mu \nu} + {\mathfrak h} U^\mu U^\nu 
+ T_{\rm EM}^{\mu \nu},
\end{equation}
where $p$, $\mathfrak h$, and $U^\mu$ are the proper pressure, the proper enthalpy density,
and the 4-velocity of the plasma, respectively.
The energy momentum conservation law is given by
\begin{equation}
\nabla_\mu T^{\mu \nu} = 0.
\label{eqenmo}
\end{equation}
The force-free condition is 
\begin{equation}
J^\mu F_{\mu \nu} = 0 ,
\end{equation}
and the general relativistic Ohm's law is
\begin{equation}
F_{\mu \nu} U^\nu = \eta \left [ J_\mu + (U^\nu J_\nu) U_\mu \right ],
\label{ohmlaw}
\end{equation}
where $\eta$ is the resistivity of the plasma.
The ideal MHD condition is given by setting $\eta=0$, $F_{\mu \nu} U^\nu = 0$.

Here, we introduce a local coordinate frame, so called 
``fiducial observer (FIDO) frame'', $(\hat{x}^0, \hat{x}^1, \hat{x}^2, \hat{x}^3)$.
Using the local coordinates of the frame $\hat{x}^\mu$, the line element becomes 
\[
ds^2 = \eta_{\mu \nu} d \hat{x}^\mu d \hat{x}^\nu 
= - d \hat{t}^2 + \sum_{i=1}^3 (d \hat{x}^i)^2,
\]
where $\eta_{\mu \nu}$ is the metric of Minkowski space-time.
Comparing this metric with Eq. (\ref{eqlinelement}), we get
\begin{equation}
d \hat{t} = \alpha dt, \verb!    !
d \hat{x}^i = h_i dx^i - \alpha \beta^i dt ,
\end{equation}
and we have partial derivative relations,
\begin{equation}
\frac{\partial}{\partial \hat{t}} = \frac{\partial t}{\partial \hat{t}}
\frac{\partial}{\partial t} 
+ \sum_i \frac{\partial x^i}{\partial \hat{t}} \frac{\partial}{\partial x^i}
= \frac{1}{\alpha} \frac{\partial}{\partial t}
+ \sum_i \frac{\beta^i}{h_i} \frac{\partial}{\partial x^i}, \verb!    !
\frac{\partial}{\partial \hat{x}^i}  =  
\frac{1}{h_i} \frac{\partial}{\partial x^i}  .
\end{equation}
Then, a contravariant vector $\hat{a}^\mu$ in the FIDO frame of an arbitrary 
contravariant vector $a^\mu$ in the global coordinates $x^\mu$ is written as,
\begin{equation}
\hat{a}^0 = \alpha a^0, \verb!   !
\hat{a}^i = h_i a^i - \alpha \beta^i a^0
\end{equation}
and the covariant vector $\hat{a}_\mu$ is
\begin{equation}
\hat{a}_0 = 
\frac{1}{\alpha} a_0 + \sum_i \frac{\beta^i}{h_i} a_i, \verb!   !
\hat{a}_i  =  \frac{1}{h_i} a_i  .
\label{covarianttransform}
\end{equation}
We use the quantities observed by the FIDO frame because they can be treated
intuitively and yield formulae more easily.
This is because the relations between the variables in the FIDO
frame are the same as these in the special theory of relativity and
similar to the Newtonian relation.

Using the quantities of electromagnetic field in the FIDO frame, 
Maxwell equations are written by the following 3+1 formalism,
\begin{equation}
\frac{\partial \hat{B}^i}{\partial t} = - \sum _{j,k} \frac{h_i}{h_1 h_2 h_3}
\epsilon ^{ijk} \frac{\partial}{\partial x^j}
\left [ \alpha h_k (\hat{E}_k - \sum _{l,m} \epsilon ^{klm} \beta ^l \hat{B}^m)
\right ],
\label{cmfa}
\end{equation}
\begin{equation}
\alpha \left ( \hat{J}^i + \hat{\rho} _{\rm e} \beta ^i \right ) 
+ \frac{\partial \hat{E}_i}{\partial t} =
\sum _{j,k} \frac{h_i}{h_1 h_2 h_3} \epsilon ^{ijk}
\frac{\partial}{\partial x^j} \left [
\alpha h_k \left ( \hat{B}^k + \sum _{l,m} \epsilon ^{klm} 
\beta ^l \hat{E}_m \right ) \right ]   ,
\label{cmam}
\end{equation}
\begin{equation}
 \sum _{i} \frac{1}{h_1 h_2 h_3} \frac{\partial}{\partial x^i}
\left ( \frac{h_1 h_2 h_3}{h_i} \hat{B}^i
\right ) = 0    ,
\label{divb0}
\end{equation}
\begin{equation}
\rho _{\rm e} = \sum _{i} \frac{1}{h_1 h_2 h_3}
\frac{\partial}{\partial x^i} \left (
\frac{h_1 h_2 h_3}{h_i} \hat{E}_i
\right )    ,
\label{dive}
\end{equation}
where $\epsilon^{ijk}=\epsilon^{0ijk}$.

For the convenience, we introduce the derivatives of arbitrary a three-vector field
$\hat{\VEC{a}}$ and an arbitrary scalar field 
$\hat{\phi}$ measured by the FIDO frame as
\begin{equation}
\nabla \cdot \hat{\VEC{a}} = \sum _i \frac{1}{h_1 h_2 h_3}
\frac{\partial}{\partial x^i} \left ( 
\frac{h_1 h_2 h_3}{h_i} \hat{a}^i
\right )  ,
\end{equation}
\begin{equation}
(\nabla \hat{\phi})_i = 
\frac{1}{h_i} \frac{\partial \hat{\phi}}{\partial x^i}  ,
\end{equation}
\begin{equation}
(\nabla \times \hat{\VEC{a}})_i = \sum _{j,k} \frac{h_i}{h_1 h_2 h_3}
\epsilon ^{ijk} \frac{\partial}{\partial x^j} (h_k \hat{a}^k)  .
\end{equation}
We express Maxwell equations in vector forms as,
\begin{equation}
\frac{\partial \hat{\bf B}}{\partial t} = -
\nabla \times [\alpha ( \hat{\bf E}-
\VEC{\beta} \times \hat{\bf B} ) ]   ,
\label{vecfa}
\end{equation}
\begin{equation}
\alpha \left ( \hat{\bf J} + \hat{\rho}_{\rm e} \VEC{\beta} \right )
+ \frac{\partial \hat{\bf E}}{\partial t}
= \nabla \times [\alpha (\hat{\bf B} +
\VEC{\beta} \times \hat{\bf E}) ]   ,
\label{vecam}
\end{equation}
\begin{equation}
\nabla \cdot \hat{\bf B} = 0   ,
\label{vecnm}
\end{equation}
\begin{equation}
\hat{\rho}_{\rm e} = \nabla \cdot \hat{\bf E} ,
\label{vecec}
\end{equation}
where $\VEC{\beta} = (\beta _1, \beta _2, \beta _3)$,
$\hat{\VEC{E}}=(\hat{E}_1, \hat{E}_2, \hat{E}_3)$,
$\hat{\VEC{B}}=(\hat{B}^1, \hat{B}^2, \hat{B}^3)$, and
$\hat{\VEC{J}}=(\hat{J}^1, \hat{J}^2, \hat{J}^3)$.\\
The 3+1 form of the force-free condition is 
\begin{equation}
\hat{\VEC{J}} \cdot \hat{\VEC{E}} = 0  , \verb!   !
\hat{\rho}_{\rm e} \hat{\VEC{E}} + 
\hat{\VEC{J}} \times \hat{\VEC{B}} = \VEC{0}  ,
\end{equation}
and Ohm's law is written by
\begin{equation}
\hat{\bf E} + \hat{\bf v} \times \hat{\bf B} = \frac{1}{\hat{\gamma}} \eta
\left [ \hat{\VEC{J}} - \rho'_{\rm e} \hat{\gamma} \hat{\VEC{v}} \right ]   ,
\label{vecimhd}
\end{equation}
where $\hat{\gamma} = \hat{U}^0$ is the Lorentz factor, 
$\hat{\VEC{v}} = (\hat{U}^1/\hat{\gamma}, \hat{U}^2/\hat{\gamma}, \hat{U}^3/\hat{\gamma})$
is the 3-velocity, and $\rho'_{\rm e} = - J^\nu U_\nu$ is the electric charge density
observed by the plasma-rest frame (the proper electric charge density).
The conservation equation of the electric charge is derived by
Eqs. (\ref{vecam}) and (\ref{vecec}) as
\begin{equation}
\frac{\partial \hat{\rho}_{\rm e}}{\partial t} +
\nabla \cdot \left [ \alpha \left ( \hat{\bf J} 
+ \hat{\rho_{\rm e}} \VEC{\beta} \right ) \right ] = 0.
\label{vecformchrgcon}
\end{equation}

We present the equations of energy and
angular momentum conservation around a spinning black hole.
When $\xi^\mu$ is a Killing vector, we have an
conservation law associated with Eq. (\ref{eqenmo})
\begin{equation}
\frac{1}{\root \of {-\|g\|}} \frac{\partial}{\partial x^\mu}
({\root \of {-\|g\|}} T^{\mu \nu} \xi _\nu)=0   .
\end{equation}
Because of $\|g\| = -(\alpha h_1 h_2 h_3)^2$, this equation yields
\begin{equation}
\frac{\partial}{\partial t} (\alpha T^{0 \nu} \xi _\nu)
= - \frac{1}{h_1 h_2 h_3} \sum _i \frac{\partial}{\partial x^i}
(\alpha h_1 h_2 h_3 T^{i \nu} \xi _\nu)  .
\end{equation}
Using the Killing vector $\chi ^\nu =(-1, 0, 0, 0)$, we have the conservation law of energy
\begin{equation}
\frac{\partial e^\infty}{\partial t}
= - \hat{\nabla} \cdot {\bf S}  ,
\label{coen}
\end{equation}
where $e^\infty \equiv \alpha T^{0 \nu} \chi _\nu$ is called
{\it energy-at-infinity} (or just ``energy'') density and ${S^i} \equiv 
\alpha h_i T^{i \nu} \chi _\nu$ is the $i$-th component of energy flux density.
Here, we also express these quantities in the FIDO frame as
\begin{equation}
e^\infty = \alpha ( \epsilon + \rho \hat{\gamma}) + \sum _i \alpha \beta^i \hat{Q}^i  ,
\label{apeai}
\end{equation}
\begin{equation}
S^i = \alpha \left [ \alpha \hat{Q}^i +  e^\infty \beta^i +
\sum _j \alpha \beta ^j \hat{T}^{ij} \right ] ,
\end{equation}
where $\epsilon + \rho \hat{\gamma} = \hat{T}^{00}$, 
$\hat{\gamma} = \hat{U}^0$, and $\hat{Q}^i = \hat{T}^{0i}$.

We separate these quantities into the hydrodynamic and electromagnetic components:
\begin{equation}
e^\infty = e^\infty _{\rm hyd} + e^\infty _{\rm EM}   ,
\end{equation}
\begin{equation}
S^i = S^i _{\rm hyd} + S^i _{\rm EM}   ,
\end{equation}
where 
\begin{equation}
e^\infty _{\rm hyd} = \alpha ( {\mathfrak h} \gamma ^2 - p)
+ \sum _i \alpha \beta^i {\mathfrak h} \gamma ^2 \hat{v}^i  ,
\label{eaikin}
\end{equation}
\begin{equation}
e^\infty _{\rm EM} = \alpha \left ( \frac{(\hat{B})^2}{2} +
\frac{\hat{(E)^2}}{2} \right ) + \sum _i \alpha \beta^i
(\hat{\bf E} \times \hat{\bf B})_i  ,
\label{eaiem}
\end{equation}
\begin{equation}
S^i _{\rm hyd} = \alpha ^2 {\mathfrak h} \gamma ^2 
\left ( 1 + \sum _j \beta ^j \hat{v}^j \right )
(\hat{v}^i + \beta ^i)  ,
\end{equation}
\begin{equation}
S^i _{\rm EM} = \alpha ^2 \left [ (\hat{\bf E} - \VEC{\beta} \times
\hat{\bf B} ) \times \left ( \hat{\bf B} + 
\VEC{\beta} \times \hat{\bf E} \right ) \right ]^i  ,
\end{equation}
where the subscripts `hyd' and `EM' indicate hydrodynamic and
electromagnetic components, respectively, and
$(\hat{E})^2 = (\hat{E}_1)^2+(\hat{E}_2)^2+(\hat{E}_3)^2$, $(\hat{B})^2 = (\hat{B}^1)^2+(\hat{B}^2)^2+(\hat{B}^3)^2$.
%
Here, ${\bf S}_{\rm EM}=(S_{\rm EM}^1,S_{\rm EM}^2,S_{\rm EM}^3)$ 
can be regarded as the Poynting vector.

The general relativistic Maxwell equations
(\ref{vecfa})-(\ref{vecec}) yield
\begin{equation}
\frac{\partial e^\infty _{\rm EM}}{\partial t}
= - \hat{\nabla} \cdot {\bf S}_{\rm EM} - 
\alpha (\hat{\bf v} + \VEC{\beta}) \cdot {\bf f}_{\rm L}  ,
\label{tren}
\end{equation}
where ${\bf f}_{\rm L} = \hat{\rho} _{\rm e} \hat{\bf E}
+ \hat{\bf J} \times \hat{\bf B}$ is the Lorentz force density.

If $\eta ^\mu = (0, 0, 0, 1)$ is the Killing vector for the azimuthal direction, we have the
equation of angular momentum conservation 
\begin{equation}
\frac{\partial l}{\partial t} = - \hat{\nabla} \cdot
{\bf M}  ,
\label{coam}
\end{equation}
where $l \equiv \alpha T^{0 \nu} \eta _\nu$
and $M^i \equiv \alpha h_i T^{i \nu} \eta _\nu$
are the total angular momentum density and the angular momentum
flux density, respectively.
Using the quantities measured in the FIDO frame,
we have 
\begin{equation}
l = h_3 \hat{Q}^3 ,
\label{apam}
\end{equation}
\begin{equation}
M^i = \alpha h_3 (\hat{T}^{i3}+ \beta ^i \hat{Q}^3). 
\end{equation}
These variables also can be divided into
the hydrodynamic and electromagnetic components, denoted by the
subscripts `hyd' and `EM', as follows:
\begin{equation}
l=l_{\rm hyd} + l_{\rm EM}  ,
\end{equation}
\begin{equation}
M^i=M^i_{\rm hyd} + M^i_{\rm EM}  ,
\end{equation}
where
\begin{equation}
l_{\rm hyd} = h_3 {\mathfrak h} \gamma ^2 \hat{v}^3  ,
\label{amkin}
\end{equation}
\begin{equation}
l_{\rm EM} = h_3 (\hat{\bf E} \times \hat{\bf B})_3  ,
\label{amem}
\end{equation}
\begin{equation}
M^i_{\rm hyd} = \alpha h_3 \left [ p \delta ^{i3} + 
{\mathfrak h} \gamma ^2 \hat{v}^i \hat{v}^3 + 
c \beta ^i {\mathfrak h} \gamma ^2 \hat{v}^3
\right ]  ,
\end{equation}
\begin{equation}
M^i_{\rm EM} = 
\alpha h_3 \left [ \left ( \frac{(\hat{B})^2}{2} + \frac{(\hat{E})^2}{2} \right ) \delta ^{i3} 
- \hat{B}^i \hat{B}^3 - \hat{E}_i \hat{E}_3
+  \beta ^i (\hat{\bf E} \times \hat{\bf B})_3 
\right ]  .
\end{equation}
In this case, from Eqs. (\ref{apeai}) and (\ref{apam}), we have a relation
of the energy and the angular momentum,
\begin{equation}
e^\infty = \alpha (\epsilon + \rho \hat{\gamma}) + \omega^3 l
= \alpha \left [ \epsilon + \rho \hat{\gamma} + \frac{\beta^3}{h_3} l  \right ],
\label{releaianm}
\end{equation}
when $\omega_1 = \omega_2=0$.
Furthermore, Eqs (\ref{eaikin}), (\ref{eaiem}), (\ref{amkin}), (\ref{amem}) yield
\begin{equation}
e^\infty _{\rm hyd} = \alpha ( {\mathfrak h} \gamma ^2 - p) + \omega^3 l_{\rm hyd},
\label{eaikinang}
\end{equation}
\begin{equation}
e^\infty _{\rm EM} = \alpha \left ( \frac{(\hat{B})^2}{2} +
\frac{\hat{(E)^2}}{2} \right ) + \omega^3 l_{\rm EM}.
\label{eaiemang}
\end{equation}
The general relativistic Maxwell equations
(\ref{vecfa})-(\ref{vecec}) read
\begin{equation}
\frac{\partial l _{\rm EM}}{\partial t}
= - \hat{\nabla} \cdot {\bf M}_{\rm EM} -h_3 f_{\rm L}^3  .
\label{tram}
\end{equation}


From now on, we consider the electromagnetic energy transport when we have the
relation between the electric field and magnetic field as
\begin{equation}
\hat{\VEC{E}} = - \hat{\VEC{v}}_{\rm F} \times \hat{\VEC{B}}.
\label{eandb}
\end{equation}
Here, $\hat{\VEC{v}}_{\rm F}$ is a certain vector field and does not always 
mean the real velocity, while in the ideal MHD case, it is identified by the plasma
velocity $\hat{\VEC{v}}$.
It is noted that the drift velocity due to the electric field $\VEC{E}$,
$\displaystyle \hat{\VEC{v}}_{E} = \frac{\hat{\VEC{E}} \times \hat{\VEC{B}}}{\hat{B}^2}$
can be used as one of vector of $\hat{\VEC{v}}_{\rm F}$.
Intuitively, $\hat{\VEC{v}}_{\rm F}$ is regarded as the velocity of the magnetic 
field lines,
while this intuition is not rigorous because we can not identify the magnetic field lines
at the different times. However, we have not so serious contradiction 
with the interpretation and we often recognize $\hat{\VEC{v}}_{\rm F}$ as the velocity
of the field line implicitly.
Using Eq. (\ref{eandb}), we have
\begin{eqnarray}
e_{\rm EM}^\infty = \alpha \left ( \frac{(\hat{B})^2}{2} + \frac{(\hat{E})^2}{2} \right ) + 
\alpha \hat{\VEC{\beta}} \cdot \left ( \hat{\VEC{E}} \times \hat{\VEC{B}} \right ) =
\alpha \left [
\frac{1}{2} (1 + \hat{v}_{\rm F \perp}^2) + \VEC{\beta} \cdot \hat{\VEC{v}}_{\rm F \perp} 
\right ] (\hat{B})^2,
\label{eeminf}
\end{eqnarray}
where $\hat{\VEC{v}}_{\rm F \perp}$ is the component of $\hat{\VEC{v}}_{\rm F}$ 
perpendicular to the magnetic field $\hat{\VEC{B}}$,
$\hat{\VEC{v}}_{\rm F} = \hat{\VEC{v}}_{\rm F \parallel} + \hat{\VEC{v}}_{\rm F \perp}$, 
$\hat{\VEC{v}}_{\rm F \parallel} \parallel \hat{\VEC{B}}$, 
$\hat{\VEC{v}}_{\rm F \perp} \perp \hat{\VEC{B}}$.
Here, we used the relations, $(\hat{E})^2 = (\hat{B})^2 \hat{v}_{\rm F \perp}^2$, 
$\hat{\VEC{E}} \times \hat{\VEC{B}} = (\hat{B})^2 \hat{\VEC{v}}_{\rm F \perp}$.
With respect to the energy transport flux density, we have
\begin{equation}
\VEC{S}_{\rm EM} = \alpha^2 \left  [ \left  \{ \frac{1}{2} (1 + \hat{v}_{\rm F \perp}^2)
+ \VEC{\beta} \cdot \hat{\VEC{v}}_{\rm F \perp} \right  \} 
(\hat{B})^2 (\hat{\VEC{v}}_{\rm F \perp} + \VEC{\beta})
+ (1 - v_{\rm F \perp}^2) \left \{ \frac{(\hat{B})^2}{2} (\hat{\VEC{v}}_{\rm F \perp} 
+ \VEC{\beta})
- (\VEC{\beta} \cdot \hat{\VEC{B}}) \hat{\VEC{B}} \right \} \right ] .
\label{semeq}
\end{equation}
Using Eqs. (\ref{eeminf}) and (\ref{semeq}), we obtain
\begin{equation}
\VEC{S}_{\rm EM} = \alpha e_{\rm EM}^\infty (\hat{\VEC{v}}_{\rm F \perp} + \VEC{\beta})
+ \alpha^2 (1 - \hat{v}_{\rm F \perp}^2) 
\left \{ \frac{\hat{B}^2}{2} (\hat{\VEC{v}}_{\rm F \perp} + \VEC{\beta})
- (\VEC{\beta} \cdot \hat{\VEC{B}}) \hat{\VEC{B}}
 \right \} .
\label{frm4sem}
\end{equation}

With respect to the angular momentum of the electromagnetic field,
assuming Eq. (\ref{eandb}), we have
\begin{eqnarray}
& l_{\rm EM} = h_3 (\hat{B})^2 \hat{v}_{\rm F \perp}^3,
\label{angmome2andb} \\
& M_{\rm EM}^i = \alpha h_3 \left [ \frac{1}{2} (1 + \hat{v}_{\rm F \perp}^2)
(\hat{B})^2 \delta^{i3} + \beta^i (\hat{B})^2 \hat{v}_{\rm F \perp}^3
-\hat{B}^i \hat{B}^3 -\hat{E}^i \hat{E}^3 \right ] .
\end{eqnarray}
In this case, we also have
\begin{equation}
e_{\rm EM}^\infty = \alpha 
\frac{(\hat{B})^2}{2} (1 + \hat{v}_{\rm F \perp}^2) + \omega^3 l_{\rm EM}.
\label{eeminfang}
\end{equation}

\section{Causal energy extraction from black holes with several kinds of electromagnetic fields
\label{sec3}}
\subsection{Force-free electromagnetic field case: Blandford-Znajek mechanism
\label{ffefbz}}

In this subsection, we consider the energy transport near the horizon in the
force-free limit case, which is assumed in the original work of Blandford-Znajek mechanism 
\citep{blandford77}.
Here, we use the Kerr metric for space-time  $(x^0, x^1, x^2, x^3)=(t, r, \theta, \phi)$
with $\omega^\phi \ge 0$ in this section.
The condition of force-free, $J^\mu F_{\mu \nu} = 0$, reads
\begin{eqnarray}
\hat{\VEC{J}} \cdot \hat{\VEC{E}} &=& 0, \label{ffcone}\\
\hat{\rho}_{\rm e} \hat{\VEC{E}} + \hat{\VEC{J}} \times \hat{\VEC{B}}
&=& \VEC{0}. \label{ffconj}
\end{eqnarray}
This means no energy and momentum transforms between the electromagnetic field
and plasma. In such a case, we can write the electromagnetic field by Eq.
(\ref{eandb}).
This is because when $\hat{\rho}_{\rm e} \neq 0$, we have
$\displaystyle \hat{\VEC{E}} = - \frac{\hat{\VEC{J}}}{\hat{\rho}_{\rm e}} \times \hat{\VEC{B}}$,
and confirm Eq. (\ref{eandb}) with 
$\displaystyle \hat{\VEC{v}}_{\rm F} = \frac{\hat{\VEC{J}}}{\hat{\rho}_{\rm e}}$.
When $\hat{\rho}_{\rm e} = 0$, we have $\hat{\VEC{J}} \times \hat{\VEC{B}} = \VEC{0}$,
that is $\hat{\VEC{J}} \parallel \hat{\VEC{B}}$. Furthermore, because of Eq. (\ref{ffcone}),
we have $\hat{\VEC{J}} \perp \hat{\VEC{E}}$, and then $\hat{\VEC{E}} \perp \hat{\VEC{B}}$.
We confirm Eq. (\ref{eandb}) with $\displaystyle \hat{\VEC{v}}_{\rm F} = \frac{1}{\hat{B}^2} 
\hat{\VEC{E}} \times \hat{\VEC{B}}$.

\begin{figure}
\includegraphics[width=7cm]{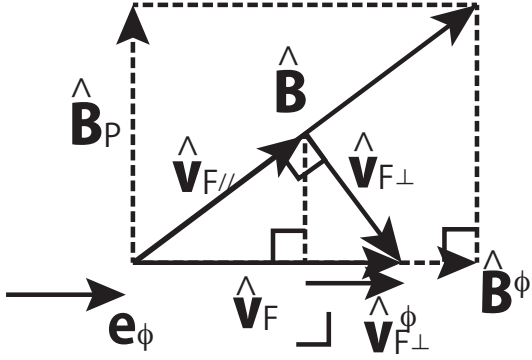}
\caption{
Geometric relation between the vectors $\hat{\VEC{B}}$, $\hat{\VEC{B}}_{\rm P}$, 
$\hat{\VEC{B}}^\phi$, $\hat{\VEC{v}}_{\rm F}$, $\hat{\VEC{v}}_{\rm F \parallel}$,
$\hat{\VEC{v}}_{\rm F \perp}$, and $\hat{\VEC{v}}_{\rm F \perp}^\phi$.
\label{bzvfperp}}
\end{figure}

In the steady-state and axisymmetry case, 
Eqs. (\ref{cmfa}), (\ref{divb0}), (\ref{ffcone}), and (\ref{ffconj}) yield
\begin{equation}
\hat{\VEC{v}}_{\rm F} = \frac{h_\phi}{\alpha} (\Omega _{\rm F}- \omega_{\phi}) \VEC{e}_\phi 
= \frac{R}{\alpha} (\Omega _{\rm F}- \omega_{\phi}) \VEC{e}_\phi ,
\label{vhf2hz}
\end{equation}
where $\Omega_{\rm F}$ is a constant along the magnetic flux surface, 
$R \equiv h_\phi = h_3$ corresponds to the distance from the $z$ axis, 
$\VEC{e}_\phi$ is
the unit vector for azimuthal direction \citep{blandford77}.
Because the triangle of $\hat{\VEC{v}}_{\rm F \perp}$ and $\hat{\VEC{v}}_{\rm F}$ 
and the triangle of $\hat{\VEC{B}}_{\rm P}$ and $\hat{\VEC{B}}$ are similar
(Fig. \ref{bzvfperp})，we found the following relation,
\begin{equation}
\frac{\hat{v}_{\rm F \perp}}{\hat{v}_{\rm F}}= \frac{\hat{B}_{\rm P}}{\hat{B}}.
\end{equation}
Here, we define $\hat{\VEC{B}}_{\rm P}$ and $\hat{\VEC{B}}_\phi$ as the
poloidal and azimuthal components of magnetic field $\hat{\VEC{B}}$, respectively.
Then, we have
\begin{equation}
\hat{v}_{\rm F \perp} = \frac{\hat{v}_{\rm F}}{\sqrt{1 + (\hat{B}^\phi/\hat{B}_{\rm P})^2}}.
\end{equation}
The Znajek boundary condition at the horizon \cite{znajek77} is expressed as 
\begin{equation}
\frac{\hat{B}^\phi}{\hat{B}_{\rm P}} = \hat{v}_{\rm F}^\phi .
\label{znajekcond}
\end{equation}
Then, very near the horizon, we also have
\begin{equation}
\hat{v}_{\rm F \perp} \approx \frac{\hat{v}_{\rm F}}{\sqrt{1 + \hat{v}_{\rm F}^2}}  ,
\end{equation}
where ``$\approx$" means asymptotic equivalence.
In the limit toward the horizon ($r \rightarrow r_{\rm H}$, $r_{\rm H}$ is
the radius of the black hole), we have
$\hat{v}_{\rm F} \rightarrow \infty$ when $\Omega_{\rm F} \neq \Omega_{\rm H}$, and
then we found $\hat{v}_{\rm F \perp} \rightarrow 1$.
Here, we write the value of $\omega^\phi$ at the horizon by $\Omega_{\rm H}$.
Eventually, using Eq. (\ref{frm4sem}) we obtain very near the horizon,
\begin{equation}
\VEC{S}_{\rm EM} = \alpha e_{\rm EM}^\infty (\hat{\VEC{v}}_{\rm F\perp} + \hat{\VEC{\beta}}) .
\end{equation}
The directions of $\hat{\VEC{v}}_{\rm F}$ of the cases of 
$\Omega_{\rm F} < \Omega_{\rm H}$ and $\Omega_{\rm F} > \Omega_{\rm H}$
are opposite because of Eq. (\ref{vhf2hz}), and the slope of the magnetic field lines
in the two cases are also opposite (Fig. \ref{bh_horizon}). 
Then, the direction of $\hat{\VEC{v}}_{\rm F \perp}$ is always directed 
toward the black hole inner region when $\Omega_{\rm F} \neq \Omega_{\rm H}$.
Then, when $e_{\rm EM}^\infty < 0$, the electromagnetic energy flux is directed outward
and the energy of the black hole is extracted through the horizon.

Next, we determine the condition of the negative energy $e_{\rm EM}^\infty < 0$
at the horizon.
When $\Omega_{\rm F} \neq \Omega_{\rm H}$,
$\VEC{v}_{\rm F \perp}$ is directed toward the black hole horizon
in both cases of $\Omega_{\rm F} < \Omega_{\rm H}$ and $\Omega_{\rm F} > \Omega_{\rm H}$.
Because the triangle of $\hat{\VEC{v}}_{\rm F \perp}$, $\hat{\VEC{v}}_{\rm F}$ 
and the triangle of $\hat{\VEC{B}}_{\rm P}$, $\hat{\VEC{B}}$ are similar
(Fig. \ref{bzvfperp}),
we found $\displaystyle \frac{\hat{v}_{\rm F \perp}^\phi}{\hat{v}_{\rm F \perp}}
=\frac{\hat{B}_{\rm P}}{\hat{B}}$, and then we obtain
\begin{equation}
\hat{v}_{\rm F \perp}^\phi  = \left ( \frac{\hat{B}_{\rm P}}{\hat{B}} \right )^2 \hat{v}_{\rm F}
= \frac{\hat{v}_{\rm F}}{1 + \hat{v}_{\rm F}^2} .
\end{equation}
Finally, using the second equation of Eq. (\ref{diff_alpbet}) and Eq. (\ref{vhf2hz}) we get
\begin{equation}
e_{\rm EM}^\infty = \left [ \frac{1}{2} 
\left ( 1 + \frac{\hat{v}_{\rm F}^2}{1 + \hat{v}_{\rm F}^2} \right )
+ \frac{\hat{\beta}^\phi \hat{v}_{\rm F}^\phi}{1 + \hat{v}_{\rm F}^2} \right ] \alpha \hat{B}^2 \\
= \frac{1}{2} \frac{\alpha^2 
+ 2 R^2 \Omega_{\rm F} (\Omega_{\rm F} - \hat{\omega}_\phi)}{\alpha^2 + R^2 (\Omega_{\rm F} 
- \hat{\omega}_\phi)^2}  \alpha \hat{B}^2.
\end{equation}
At the horizon, ($\alpha \longrightarrow 0$，
$\omega_\phi \longrightarrow \Omega_{\rm H}$), 
Eqs. (\ref{vhf2hz}) and  (\ref{znajekcond}) yield 
$\displaystyle \hat{B} = \sqrt{(\hat{B}^\phi)^2+(\hat{B}_{\rm P})^2}
\approx |\hat{B}^\phi| = \left | \frac{R}{\alpha} (\Omega_{\rm F} - \Omega_{\rm H}) \right |
\hat{B}_{\rm PH}$ because $| \hat{B}^\phi | \gg \hat{B}_{\rm P}$,
where $\hat{B}_{\rm PH}$ is the value of $\hat{B}_{\rm P}$ at the horizon. 
Eventually, at the horizon, we found 
\begin{eqnarray}
& \displaystyle e_{\rm EM}^\infty \approx \frac{R_{\rm H}^2}{\alpha} 
\Omega_{\rm F} (\Omega_{\rm F} - \Omega_{\rm H}) (\hat{B}_{\rm PH})^2, 
\label{form2eemi} \\
& \VEC{S}_{\rm EM}  = R_{\rm H}^2 \Omega_{\rm F} (\Omega_{\rm F} - \Omega_{\rm H})
(\hat{B}_{\rm PH})^2 (\hat{\VEC{v}}_{\rm F \perp} + \VEC{\beta}),
\label{form2sem}
\end{eqnarray}
where $R_{\rm H}$ is the value of $R = h_\phi$ at the horizon.
It is noted that the radial component of the electromagnetic energy flux is
identical to the simple equation given by \citet{mckinney04} (Eq. (34) in the paper),
if we set the force-free condition at the horizon, $\hat{\VEC{v}}_{\rm F}^\perp
= \VEC{e}_r$.
Then, when $0 < \Omega_{\rm F} < \Omega_{\rm H}$, the negative energy of the 
electromagnetic field is realized ($e_{\rm EM}^\infty < 0$) and
the rotation energy of the black hole is extracted.
This is exactly the same condition of the Blandford-Znajek mechanism. This suggests
even in the Blandford-Znajek mechanism, to extract the black hole rotational energy,
the negative energy of the electromagnetic field is utilized as a mediator.
In conclusion, putting the negative electromagnetic energy into the black hole,
the black hole rotational energy is extracted causally in the Blandford-Znajek mechanism.

Sometimes the energy extraction of the rotating black hole is intuitively explained
by the torque of magnetic field at the horizon. This intuitive explanation is not
appropriate with respect causality. Because at the horizon no torque affects
the matter and field outside of the horizon from these inside of the horizon.
Eqs. (\ref{form2eemi}) and (\ref{form2sem}) suggest that the falling-down 
of the negative (electromagnetic) energy into the black hole could
decrease the black hole energy to extract the black hole energy.

\subsection{Ideal MHD case: MHD Blandford-Znajek mechanism/ MHD Penrose process
\label{mhdbz}}

\begin{figure}
\includegraphics[width=8cm]{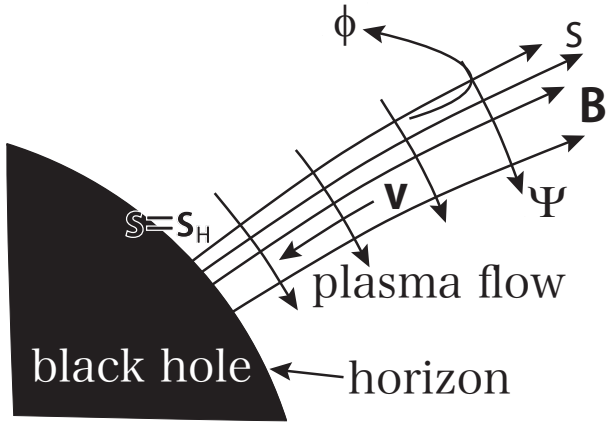}
\caption{
Magnetic field surfaces in the steady-sate ideal MHD case and
the spatially orthogonal coordinates $(s, \Psi, \phi)$.
\label{idealmhdcase}}
\end{figure}

We consider the ideal MHD case in the space-time around the spinning black hole.
We assume the situation is stationary and axisymmetric as the same as
the force-free case in the previous section.
In such a case, the magnetic flux surfaces are stationary and axisymmetric 
and are expressed as constant azimuthal component of vector potential, $A_\phi$.
We introduce the new coordinate system $(t, s, \Psi, \phi)$, where 
$t$ is the time of Kerr space-time, $\phi$ is 
the azimuthal coordinate, $\Psi=A_\phi$, and
the coordinate $s$ is set outwardly along the intersection line of a magnetic surface
and the meridian plane ($\phi$= const.)(Fig. \ref{idealmhdcase}).
Here, we set the coordinates $s$ so that it is perpendicular to the coordinate
$\Psi$. The $s$ coordinate at the horizon is $s_{\rm H}$. 
Essentially, this coordinate system corresponds to the Boyer-Linquist
coordinate $(t, r, \theta, \phi)$ 
where $t=t$, $s=s(r,\theta)$, $\Psi=\Psi(r, \theta)$, and $\phi=\phi$.
Then, the length of a line element in the space-time of the rotating black hole
is given by
\[
ds^2 = -h_t^2 dt^2 + h_s^2 ds^2 + h_\Psi^2 d\Psi^2 + h_\phi^2 d\phi^2
- 2 h_\phi^2 \omega_\phi dt d\phi.
\]
We assume the ideal MHD condition, $U^\mu F_{\mu \nu}=0$, which yields 
\begin{equation}
\hat{\VEC{E}} + \hat{\VEC{v}} \times \hat{\VEC{B}} = \VEC{0}.
\label{idealmhdcond}
\end{equation}
Using the coordinates $(s, \Psi, \phi)$, Eqs. (\ref{cmfa})--(\ref{dive}), 
(\ref{coen}), (\ref{coam}), and (\ref{idealmhdcond}) yield 
the following conservation variables along the magnetic surface:
\begin{eqnarray} 
\dot{M}(\Psi) &  =& h_\phi h_\Psi \rho \alpha \hat{U}^s
= \frac{\alpha \rho \hat{U}^s}{\hat{B}^s}, \label{constmd} \\
B^s(\Psi) & =& h_\Psi h_\phi \hat{B}^s =1, \label{constbs} \\
\Omega_{\rm F} (\Psi) & =& \frac{\alpha}{h_\phi} \left [
\hat{v}^\phi + \beta^\phi - \frac{\hat{B}^\phi}{\hat{B}^s} \hat{v}^s \right ] , 
\label{constwf}\\
L(\Psi) & =& h_\phi \left [ \frac{\mathfrak h}{\rho} \hat{U}^\phi 
- \frac{\alpha}{\dot{M}} \hat{B}^\phi
\right ]  , \label{constl} \\
H(\Psi) & =& \frac{\mathfrak h}{\rho} 
[\alpha \hat{\gamma} - h_\phi (\Omega_{\rm F} - \omega_{\phi} )  ]
= \frac{\mathfrak h}{\rho} \alpha (\hat{\gamma} - \hat{v}^\phi_{\rm F} \hat{U}^\phi).
\label{consth}
\end{eqnarray}
It is noted that quantities with hats are variables observed by the FIDO frame.
It is also noted that the distribution of $\Psi$ is determined by the transverse
equation called the ``Grad-Shafranov equation"\citep{beskin00}.
Recently, numerical simulations of GRMHD provide the more complete feature of the 
mechanism like the distribution of Poynting flux over the event horizon, 
the relative importance of negative energy-at-infinity fluid and electromagnetic field,
the energy flux from the black hole to the disk through the magnetic field lines, etc.
\citep{mckinney12,hawley06}.
It is noted that the numerical, time-dependent simulations showed that 
magneto-rotational instability (MRI) always causes fluctuations and no steady state 
of plasma and magnetic field is found.

At the black hole horizon, the lapse function $\alpha$ becomes 0, $h_s$ becomes infinite,
while $\displaystyle \omega^\phi = \frac{\alpha \beta^\phi}{h_\phi} \longrightarrow \Omega_{\rm H}$,
$h_\Psi \longrightarrow h_{\Psi \rm H}$, $h_\phi \longrightarrow R_{\rm H}$ 
are finite except on the $z$ axis.
Hereafter, we discuss the quantities along a certain fixed magnetic flux surface $\Psi=\Psi$.

\begin{figure}
\includegraphics[width=14cm]{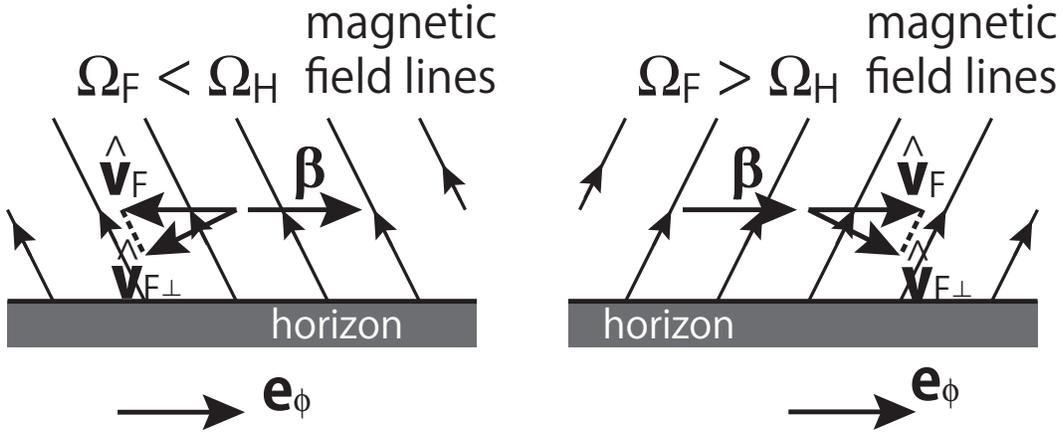}
\caption{
Geometry of vectors  $\hat{\VEC{v}}_{\rm F}$, $\hat{\VEC{v}}_{\rm F \perp}$,
$\hat{\VEC{\beta}}$ and the magnetic field lines in the case of 
$\Omega_{\rm F} < \Omega_{\rm H}$ (left panel) and $\Omega_{\rm F} > \Omega_{\rm H}$ 
(right panel) near the horizon.
\label{bh_horizon}}
\end{figure}

Because the horizon is not a real singular surface, and the density $\rho$ and
pressure $p$ are measured by the plasma rest frame, $\rho$ and $p$ should be
finite at the horizon. Then, from Eqs. (\ref{constmd}) and (\ref{constbs}),
$\alpha \hat{U}^s$ and $\hat{B}^s$ must be finite at the horizon, where we write $\hat{B}^s$ at
the horizon by $\hat{B}_{\rm H}^s$. 
At the horizon, the plasma falls vertically to the horizon at the light velocity
\footnote{This is also derived as follows. Extremely near the horizon, 
$\alpha \hat{B}^\phi$ is finite, because $\hat{v}^s$ is finite. 
Then, from Eq. (\ref{constl}), $\hat{U}^\phi$ is finite.
At the horizon, because $\alpha \hat{U}^s$ is finite and $\alpha \longrightarrow 0$, 
$\hat{U}^s$ and $\hat{\gamma}= \sqrt{1 + (\hat{U}^s)^2+(\hat{U}^\phi)^2}$ are infinite.
Then, $\displaystyle \hat{v}^\phi = \frac{\hat{U}^\phi}{\hat{\gamma}}$ 
becomes zero at the horizon and $\tilde{\gamma} = |\hat{U}^s|$.
Finally, at the horizon, $\displaystyle \hat{v}^s = \frac{\hat{U}^s}{\hat{\gamma}} = -1$.},
$\hat{v}^\phi = \hat{v}^\Psi = 0$, $\hat{v}^s=-1$, and second equation of Eq. (\ref{diff_alpbet})
and Eq. (\ref{constwf}) yield
\begin{equation}
\frac{\alpha \hat{B}^\phi}{\hat{B}^s} \approx R_{\rm H}(\Omega_{\rm F} - \Omega_{\rm H}) .
\label{bnd2hz}
\end{equation}
Using Eqs. (\ref{idealmhdcond}) and (\ref{constwf}), we have 
\begin{equation}
\hat{\VEC{E}} = - \hat{\VEC{v}} \times \hat{\VEC{B}}
= \hat{B}^s \left ( - \hat{v}^\phi + \hat{v}^s \frac{\hat{B}^\phi}{\hat{B}^s} \right )
\VEC{e}_\phi \times \VEC{e}_s
= - \frac{R}{\alpha} (\Omega_{\rm F} - \omega_\phi) \VEC{e}_\phi \times
(\hat{B}^s \VEC{e}_s) = - \hat{\VEC{v}}_{\rm F} \times \hat{\VEC{B}} ,
\end{equation}
where we put $\displaystyle \hat{\VEC{v}}_{\rm F} = \frac{R}{\alpha} 
(\Omega_{\rm F} - \omega^\phi) \VEC{e}_\phi$ and
$\VEC{e}_\phi$, $\VEC{e}_s$ are the unit base vectors along the $\phi$ and $s$
coordinates, respectively.
Very near the horizon, we have 
\begin{equation}
\hat{\VEC{v}}_{\rm F} \approx \frac{R_{\rm H}}{\alpha} 
(\Omega_{\rm F} - \Omega_{\rm H}) \VEC{e}_\phi.
\label{eqvf}
\end{equation}
Eqs. (\ref{bnd2hz}) and (\ref{eqvf}) present the geometrical disposition of vectors
$\hat{\VEC{B}}$ and $\hat{\VEC{v}}_{\rm F}$, as shown in Fig. \ref{bh_horizon}.
When $\Omega_{\rm F} \neq \Omega_{\rm H}$,
we found that
the vector of $\VEC{v}_{\rm F}$ is always directed toward the black hole inner region.

Intuitively, at the horizon of the rotating black hole,
the plasma falls into the black hole radially with the speed of light
($\hat{v}^s = -1$, $\hat{v}^\phi = 0$ at $s = s_{\rm H}$).
When the azimuthal component of magnetic field is finite outside of the horizon
and stationary, the magnetic field lines are twisted extremely strongly
near the horizon in appearance because of Eq. (\ref{constwf}) where $\alpha \hat{B}_\phi$ 
is uniform along the magnetic surface and $\alpha$ vanishes at the horizon.
This is due to difference in the lapse of time near the black hole and
is apparent feature in the Kerr metric.
In such case, the perpendicular
component of the velocity to the magnetic field is identical to the plasma
velocity and then we have 
\begin{equation}
\hat{v}_{\rm F}^\perp = \hat{v}^\perp = 1
\label{eqvfperp}
\end{equation}
at the horizon 
\footnote{This equation is also derived as follows.
Using similarity of the triangle of $\hat{\VEC{\VEC{v}}}_{\rm F \perp}$ and 
$\hat{\VEC{v}}_{\rm F}$ and the triangle of $\hat{\VEC{B}}_{\rm P}=\hat{\VEC{B}}^s$
and $\hat{\VEC{B}}$ (Fig. \ref{bzvfperp}), we found
$\displaystyle \frac{\hat{v}_{\rm F \perp}}{\hat{v}_{\rm F}}= \frac{\hat{B}_{\rm P}}{\hat{B}} 
= \frac{\hat{B}^{s}}{\hat{B}}$,
where $\displaystyle \hat{B} = \sqrt{(\hat{B}^s)^2+ (\hat{B}^\phi)^2}$ and
$\hat{B}_{\rm P} = \hat{B}^s$.
Very near the horizon, because $\hat{B}^\phi$ is much larger than $\hat{B}^s$,
we have $\displaystyle \frac{\hat{B}^{s}}{\hat{B}} \approx \frac{\hat{B}^s}{|\hat{B}^\phi|}$
and then 
$\displaystyle \frac{\hat{v}_{\rm F \perp}}{\hat{v}_{\rm F}} \approx 
\frac{\hat{B}^s}{|\hat{B}^\phi|}$.
Using Eqs. (\ref{bnd2hz}) and (\ref{eqvf}), at the horizon we confirm
$\displaystyle \hat{v}_{\rm F \perp} = \frac{\hat{B}^s}{|\hat{B}^\phi|} \hat{v}_{\rm F} =1$.}.
Then, the electromagnetic energy flux density at the horizon is given by
\begin{equation}
\VEC{S}_{\rm EM} =\alpha e_{\rm EM}^\infty (\hat{\VEC{v}}_{\rm F \perp} + \hat{\VEC{\beta}}) ,
\end{equation}
from Eq. (\ref{frm4sem}).
When $e_{\rm EM}^\infty$ becomes negative at the horizon, the electromagnetic
energy is transported outwardly through the horizon when $\Omega_{\rm F} \neq \Omega_{\rm H}$,
because $\hat{\VEC{v}}_{\rm F \perp}$ is always directed inwardly toward the black hole inner 
region (see Fig. \ref{bh_horizon}).
Here, because $\hat{\VEC{v}}_{\rm F \perp}$ vanishes if $\Omega_{\rm F} = \Omega_{\rm H}$,  
no electromagnetic output is expected, then we consider only the case of
$\Omega_{\rm F} \neq \Omega_{\rm H}$ case.

As shown in Eq. (\ref{eeminf}), the electromagnetic energy-at-infinity density is given by
\begin{equation}
e_{\rm EM}^\infty = \alpha \left [ \frac{1}{2} (1 + \hat{v}_{\rm F \perp}^2)
+ \beta^\phi \hat{v}_{\rm F \perp}^\phi \right ] (\hat{B})^2 .
\end{equation}
With Fig. \ref{bzvfperp}, we found
$\displaystyle
\frac{\hat{v}_{\rm F \perp}^\phi}{\hat{v}_{\rm F \perp}} 
= \frac{\hat{v}_{\rm F \perp}}{\hat{v}_{\rm F}^\phi}$ ,
and then we have $\displaystyle 
\hat{v}_{\rm F \perp}^\phi = \frac{(\hat{v}_{\rm F \perp})^2}{\hat{v}_{\rm F}^\phi}$.
At the horizon, using Eqs. (\ref{eqvf}) and (\ref{eqvfperp}),
we have
\begin{equation}
\hat{v}_{\rm F \perp}^\phi \approx \frac{\alpha}{R_{\rm H} (\Omega_{\rm F} - \Omega_{\rm H})}.
\end{equation}
Using Eq. (\ref{bnd2hz}), we also have 
$\displaystyle \hat{B} \approx |\hat{B}^\phi| \approx 
\left | \frac{R_{\rm H}}{\alpha} (\Omega_{\rm F} - \Omega_{\rm H}) \right | | \hat{B_{\rm H}}^s |$
at the horizon because of $| \hat{B}^\phi | \gg | \hat{B}^s |$. 
Eventually, we obtain
\begin{eqnarray} 
&\displaystyle e_{\rm EM}^\infty \approx \frac{R_{\rm H}^2}{\alpha} \Omega_{\rm F} 
(\Omega_{\rm F} - \Omega_{\rm H}) (\hat{B}_{\rm H}^s)^2, 
\label{eemhzmhd} \\
& \VEC{S}_{\rm EM} = R_{\rm H}^2 \Omega_{\rm F} (\Omega_{\rm F} - \Omega_{\rm H})
(\hat{B}_{\rm H}^s)^2 (\hat{\VEC{v}}_{\rm F \perp} + \VEC{\beta})
\label{semhzmhd}.
\end{eqnarray}
This clearly shows that when $0 < \Omega_{\rm F} < \Omega_{\rm H}$, $e_{\rm EM}^\infty$ becomes
negative and the electromagnetic energy flux directs outward through the horizon.
It is surprising that not only the condition of the electromagnetic energy extraction 
from the black hole
but also the expression of energy density and the energy flux density at the horizon are
the same as those of the Blandford-Znajek mechanism (force-free case).

In the above two cases of electromagnetic extraction of the black hole rotational 
energy, the negative electromagnetic energy-at-infinity is required
as a mediator to extract the black hole 
rotational energy through the horizon causally.  As shown in Eq. (\ref{eeminfang}), we have
$e_{\rm EM}^\infty = \alpha u_{\rm EM} + \omega^\phi l_{\rm EM}$,
where $\displaystyle u_{\rm EM} = \frac{(\hat{E})^2}{2}
+ \frac{(\hat{B})^2}{2}$ is the electromagnetic energy density in the FIDO frame. 
To realize the negative electromagnetic energy, 
the angular momentum of
the electromagnetic field $l_{\rm EM}$ should become less than $- \alpha u_{\rm EM}/\omega^\phi$.
Locally the angular momentum should be conserved because of
Eq. (\ref{coam}) and then
redistribution of the angular momentum is required. 
In the Penrose process, fission of a particle is utilized for redistribution of 
the angular momentum and production of a particle with negative energy-at-infinity.
Equation (\ref{tram}) indicates that dynamically only the magnetic force (the magnetic tension 
in the axisymmetric case) and the Lorentz force 
can redistribute the electromagnetic angular momentum.
In the ideal MHD case, magnetic tension plays an important role to redistribute
the electromagnetic angular momentum and realize the negative electromagnetic
energy. This mechanism of energy extraction with negative 
electromagnetic energy is often confused with the (original) Blandford-Znajek
mechanism, where the force-free condition is used, as we did in section \ref{sec1}.
However, strictly speaking, they should be distinguished. 
From now on, in this paper, we call the ideal MHD process 
with the negative electromagnetic energy ``MHD Blandford-Znajek mechanism".
In the MHD Blandford-Znajek mechanism, we have to take the hydrodynamic energy
flux of the plasma flow into account to discuss the net energy flux from/into
the black hole.

In fact, in the ideal MHD case, the black hole rotational energy can be also extracted
with the negative {\it hydrodynamic} energy of the plasma. The hydrodynamic energy
flux density is
\[
\VEC{S}_{\rm hyd} = \alpha (e_{\rm hyd}^\infty + \alpha p) (\hat{\VEC{v}}
+ \hat{\VEC{\beta}}) .
\]
Then, near the horizon if the plasma with $\alpha e_{\rm hyd}^\infty < 0$ falls into the black hole,
the energy is transported outwardly through the horizon because $\alpha \longrightarrow 0$ 
at the horizon.
If $\VEC{S}_{\rm EM}$ is directed outward, $\alpha e_{\rm hyd}^\infty$ must be smaller
than zero to extract the black hole rotational energy.
This extraction mechanism of black hole rotational energy is called ``MHD Penrose process''
\citep{takahashi90,hirotani92,koide02,koide03}.
The hydrodynamic energy is given by $e_{\rm hyd}^\infty = \alpha ({\mathfrak h} \gamma
-p) + \VEC{\omega} \cdot \VEC{l}_{\rm hyd}$ where $\VEC{\omega} = (\omega_1, \omega_2, \omega_3)$
and $l_{\rm hyd} = {\mathfrak h} \gamma^2 h_3 \hat{v}^3$ is the hydrodynamic angular
momentum density. To realized the negative hydrodynamic energy, 
$\displaystyle l_{\rm hyd}^3 < - \frac{\alpha ({\mathfrak h} \gamma^2 -p)}{\omega^3}$.
The angular momentum is conserved and the redistribution of hydrodynamic angular
momentum is also required. The redistribution of hydrodynamic
angular momentum is caused by the Lorentz force shown in Eq. (\ref{tram}).

To distinguish the MHD Blandford-Znajek mechanism and MHD Penrose process,
we should observe the electromagnetic and hydrodynamic energy-at-infinity density
($e_{\rm EM}^\infty$ and $e_{\rm hyd}^\infty$).
If the electromagnetic energy plays a main role to extract the black hole
energy, we recognize the process as the MHD Blandford-Znajek mechanism.
On the other hand, the hydrodynamic or plasma energy has an important role
for the extraction, it is recognized as the MHD Penrose process.
In the real cases, both of them are possible, while some long-term simulations
indicate that MHD Penrose process
is transient and the MHD Blandford-Znajek mechanism is dominant in the
later phase of the simulations \citep{komissarov05,mckinney06}.
The electromagnetic extraction mechanisms of black hole rotational energy
picked up in this paper are restricted to those
in the steady-state, axisymmetric cases. 
Recently, the long term GRMHD simulations showed 3-D dynamics of plasma
interacting with the magnetic field around the rotating black hole
\citep{mckinney12}. Strictly speaking, the results of this paper
are not applicable to the time-dependent, axiasymmetric numerical
results. The generalization of the results of this paper for such
time-dependent, axiasymmetric numerical results is required.

\subsection{Electromagnetic wave case: Superradiance \label{superradiance}}

We mention the electromagnetic wave energy transport through the horizon briefly.
We use the Kerr metric for the space-time $(x^0, x^1, x^2, x^3)=(t, r, \theta, \phi)$ 
around the spinning black hole, where we set $\omega^\phi \ge 0$.
We consider the stationary solution of the electromagnetic wave in the vacuum,
where each component of the electromagnetic field is proportional to 
$f(r,\theta) e^{-i \omega t + i m \phi}$ ($f$ is a function of $r$ and $\theta$).
We use the short wavelength limit of the electromagnetic wave,
$\displaystyle | \VEC{k} | \gg \left | \frac{1}{h_i}  
\frac{\partial}{\partial x^i} g_{\mu \nu}\right|$ 
($i=1,2,3$, $\mu, \nu= 0,1,2,3$), where $\VEC{k}$ is the wavenumber of the
electromagnetic wave in a local region, which is fixed at the global coordinates.
In a vacuum ($\hat{\VEC{J}}=\VEC{0}$，$\hat{\rho}_{\rm e}=0$),
Eqs. (\ref{vecfa})---(\ref{vecec}) 
in the FIDO frame yield
\begin{equation}
\hat{\VEC{E}}  = - \frac{\hat{\VEC{k}}}{\hat{\omega}} \times \hat{\VEC{B}} ,\verb!   !
\hat{\VEC{B}}  =  \frac{\hat{\VEC{k}}}{\hat{\omega}} \times \hat{\VEC{E}} ,
\end{equation}
where $\hat{\VEC{k}}$ and $\hat{\omega}$ are the wave number and 
angular frequency of the electromagnetic wave in the FIDO frame.
These equations read the dispersion relation, $\hat{\omega} = \pm \hat{k}$ and 
the relation $\hat{\VEC{k}} \perp \hat{\VEC{B}}$.
In this case, we identify
\begin{equation}
\hat{\VEC{v}}_{\rm F \perp} = - \frac{\hat{\VEC{k}}}{\hat{\omega}} .
\label{vfperpideal}
\end{equation}
Because of $\hat{v}_{\rm F \perp} = | \hat{\VEC{k}}/\hat{\omega} | = 1$, 
using Eq. (\ref{frm4sem}), we have
\begin{equation}
\VEC{S}_{\rm EM} = \alpha e_{\rm EM}^\infty (\hat{\VEC{n}} + \hat{\VEC{\beta}}).
\end{equation}
When electromagnetic wave passes through the horizon and enters into the black hole,
if $\alpha e_{\rm EM}^\infty$ is negative, the rotational energy of the black hole
decreases.
In this case, Eqs. (\ref{angmome2andb}), (\ref{eeminfang}), and (\ref{vfperpideal}) read
\begin{equation}
e_{\rm EM}^\infty = \alpha (\hat{B})^2  + \omega^3 l_{\rm EM}
=  \alpha \left ( 1 + \omega^3 h_3 \frac{\hat{k}^\phi}{\alpha \hat{\omega}} \right ) (\hat{B})^2  .
\end{equation}
Very near the horizon, we have $\displaystyle e_{\rm EM}^\infty \approx \omega^3 R
\frac{\hat{k}^\phi}{\hat{\omega}} ( \hat{B})^2$.
Because the 4-wavenumber $k_\mu = (- \omega, k_1, k_2, k_3)$ is the covariant vector,
using Eq. (\ref{covarianttransform}), we have 
\begin{equation}
- \hat{\omega} = \frac{1}{\alpha} (- \omega) + \frac{\beta^3}{h_3} k_3
= - \frac{1}{\alpha} (\omega - \omega_3 k_3), \verb!   !
\hat{k}_3 = \frac{1}{h_3} k_3 = \frac{m}{h_3}.
\end{equation}
Then, the energy density of the electromagnetic wave very near the horizon is
approximately given by
\begin{equation}
e_{\rm EM}^\infty \approx \Omega_{\rm H} \alpha \frac{m}{\omega - m \Omega_{\rm H}} (\hat{B})^2.
\end{equation}
When $\omega < m \Omega_{\rm H}$, the negative energy at the horizon appears and the
rotational energy of the black hole is extracted.
This extraction mechanism corresponds to the ``superradiance".
To produce the negative energy of the electromagnetic wave, the redistribution
of the angular momentum is required. To understand the redistribution process, we have to
consider the structure of the solution of the electromagnetic wave in the ergosphere.

\section{Discussion \label{sec4}}
In this paper, we showed simple formulae (Eqs. (\ref{frm4sem}) and (\ref{eeminfang})) 
to aid in building
physical intuition on the causal extraction mechanism of the black hole energy
by the electromagnetic fields with the negative electromagnetic energy
produced in the ergosphere.
In three cases of force-free, ideal MHD conditions and electromagnetic wave in vacuum, 
at the horizon we found that $\hat{v}_{\rm F \perp}=1$ and then we have 
$\VEC{S}_{\rm EM} = \alpha e_{\rm EM}^\infty (\hat{\VEC{v}}_{\rm F \perp}+\VEC{\beta})$.
To extract the black hole rotational energy causally, we have to put the
negative electromagnetic energy down into the black hole through the horizon. 
To produce the negative electromagnetic energy,
because of the angular momentum conservation (\ref{eeminfang}),
we should redistribute the angular momentum of the electromagnetic field,
where we require the negative electromagnetic angular momentum density,
\begin{equation}
l_{\rm EM} < - \frac{\alpha (\hat{B})^2}{\omega^\phi}
= - \frac{R (\hat{B})^2}{\beta^\phi}    < 0,
\end{equation}
at the horizon (see Eq. (\ref{eeminfang})).
To realize the negative angular momentum azimuthal component,
the angular momentum should be redistributed because the total
angular momentum is conserved.
The redistribution of the angular momentum of the electromagnetic field
is caused by the electromagnetic torque \citep{koide03,gammie04,hawley06,krolik05}.

This point of view is originated on the Penrose process \citep{penrose69},
which uses negative mechanical energy of a particle.
In fact, equations of the energies of matter and electromagnetic field 
have similar forms as shown in Eqs. (\ref{eaikinang}) and (\ref{eaiemang}).
With the viewpoint, in general, we classify the known mechanisms 
of energy extraction from the black hole as shown in Table \ref{clasmech}.
The Penrose process is well known and is shortly mentioned in section \ref{sec1}.
The Blandford-Znajek mechanism, MHD Blandford-Znajek mechanism, 
and the MHD Penrose process were explained in the previous sections. 
We showed that in all electromagnetic mechanisms of the energy extraction
from the spinning black hole, the negative electromagnetic energy is utilized
as a mediator for the causal energy extraction through the horizon.
We confirmed that the condition of the energy extraction is given by
the realization condition of the negative energy at the horizon.
The magnetic Penrose process was not discussed in this paper.
In the magnetic Penrose process, a particle interacts with the electromagnetic field 
and falls to the negative energy orbit. The negative energy of the particle is used to
extract the black hole rotational energy. 
This is just the Penrose process with the electromagnetic interaction instead of fission.
The superradiance was mentioned in subsection \ref{superradiance}. 
We found the electromagnetic wave
with negative energy is used to extract the black hole rotational energy.
We also add the energy extraction mechanism with the magnetic reconnection
in the ergosphere in Table \ref{clasmech} \citep{koide09}.


We discuss the coincidence of the formulae of the energy density and 
the energy flux density
of the electromagnetic field at the horizon for the force-free and MHD
Blandford-Znajek mechanisms as shown by Eqs. (\ref{form2eemi}), (\ref{form2sem}), 
and (\ref{eemhzmhd}), (\ref{semhzmhd}) in sections \ref{ffefbz} and \ref{mhdbz}, 
although the conditions of the two mechanisms are different.
On a posteriori reasoning, we have the coincident expressions of the electric
field $\hat{\VEC{E}} = - \hat{\VEC{v}}_{\rm F} \times \hat{\VEC{B}}$,
$\displaystyle \hat{\VEC{v}}_{\rm F}=\frac{R_{\rm H}}{\alpha} (\Omega_{\rm F}
- \Omega_{\rm H}) \VEC{e}_\phi$ in the assumption of stationary, axisymmetric
conditions for the both cases. Furthermore, we have the coincident
boundary condition at the horizon $\hat{v}_{\rm F \perp} \longrightarrow 1$ and 
$\displaystyle \hat{B} = \hat{B}_{\rm P} \frac{\hat{v}_{\rm F}}{\hat{v}_{\rm F \perp}}$ 
for the both cases. These leading equations for the both cases are the same
and then we have the coincident formulae for the both mechanisms.

Here, we remark on the overlap of the ideal MHD and force-free conditions.
The conditions of ideal MHD (Eq. (\ref{idealmhdcond})) and force-free 
(Eq. (\ref{ffconj})) can both be satisfied if 
$\hat{\VEC{J}} = \rho_e \hat{\VEC{v}} + \hat{\VEC{J}}_\parallel$
and $\rho_{\rm e} \neq 0$, 
where $\hat{\VEC{J}}_\parallel$ is a vector parallel to the magnetic field 
$\hat{\VEC{B}}$. The vector $\hat{\VEC{J}}_\parallel$ corresponds to the net current 
density along the magnetic field lines at the plasma rest-frame.
Alternatively, in ideal MHD simulations, the ``force-free" condition is
often defined by $\hat{B}^2/(2 \rho h) \gg 1$ even if 
$\hat{\VEC{J}} - \rho_e \hat{\VEC{v}}$ is not 
parallel to $\hat{\VEC{B}}$. 

In the astrophysical situation like AGNs, which mechanism is mostly expected to extract
the black hole rotational energy and activate the region near the black hole?
We think the MHD Blandford-Znajek mechanism is most promising process rather than
the original Blandford-Znajek mechanism. Because the plasma near the black hole
is expected to be relativistically hot, the plasma beta $\beta_{\rm p} = 2p/B^2$ 
never vanishes. Of course, the original Blandford-Znajek mechanism is applicable
as an approximation with respect to the very strong magnetic field case.
Such very low plasma beta is expected at the higher-latitude of the
black hole magnetosphere and the fast component of a relativsitc jet.

\begin{deluxetable}{lllll}
\tabletypesize{\scriptsize}
\rotate
\tablecaption{Classification of various mechanisms of energy extraction from black hole.
\label{clasmech}}
\tablewidth{0pt}
\tablehead{
\colhead{mechanism} & \colhead{form of negative energy} 
& \colhead{\parbox[c]{0.3\textwidth}{\begin{flushleft} torque for redistribution of angular momentum
\end{flushleft}}}
& \colhead{output energy} & \colhead{references}
}
\startdata
Penrose process & mechanical energy of particle & force of particle fission
& mechanical energy of particle & \citet{penrose69} \\
\parbox[c]{0.25\textwidth}{\begin{flushleft} magnetic Penrose process \end{flushleft}} 
& \parbox[c]{0.25\textwidth}{\begin{flushleft} mechanical energy of electrically charged particles 
\end{flushleft}}  & 
\parbox[c]{0.25\textwidth}{\begin{flushleft} electromagnetic force \end{flushleft}}  
& \parbox[c]{0.25\textwidth}{\begin{flushleft}  mechanical energy of electrically charged particles 
\end{flushleft}}
& \citet{wagh89} \\ 
\parbox[c]{0.25\textwidth}{\begin{flushleft} force-free Blandford-Znajek mechanism 
\end{flushleft}} & electromagnetic energy & 
\parbox[c]{0.25\textwidth}{\begin{flushleft} electromagnetic tension force (force-free)
\end{flushleft}}  
& electromagnetic energy & \citet{blandford77} \\ 
\parbox[c]{0.25\textwidth}{\begin{flushleft} MHD Blandford-Znajek mechanism
\end{flushleft}} & electromagnetic energy & 
\parbox[c]{0.25\textwidth}{\begin{flushleft} electromagnetic tension force (MHD) \end{flushleft}}  
& \parbox[c]{0.25\textwidth}{\begin{flushleft} electromagnetic energy and kinetic energy 
(Alfven wave) \end{flushleft}}
& \parbox[c]{0.25\textwidth}{\begin{flushleft} \citet{takahashi90,koide03,komissarov05}
\end{flushleft}} \\ 
MHD Penrose process & mechanical energy of plasma & 
\parbox[c]{0.25\textwidth}{\begin{flushleft} Lorentz force (magnetic tension, MHD)\end{flushleft}}  
&\parbox[c]{0.25\textwidth}{\begin{flushleft} electromagnetic energy and kinetic energy (Alfven wave)
\end{flushleft}} &  
\parbox[c]{0.25\textwidth}{\begin{flushleft} \citet{takahashi90,hirotani92,koide02,koide03} 
\end{flushleft}} \\
\parbox[c]{0.25\textwidth}{\begin{flushleft} energy extraction with magnetic reconnection 
\end{flushleft}} & 
mechanical energy of plasmoid & 
\parbox[c]{0.25\textwidth}{\begin{flushleft} magnetic tension due to magnetic reconnection 
\end{flushleft}} 
& mechanical energy of plasmoid & \citet{koide09} \\ 
superradiance & 
\parbox[c]{0.25\textwidth}{\begin{flushleft} electromagnetic energy of electromagnetic wave 
\end{flushleft}} & \parbox[c]{0.25\textwidth}{\begin{center} 
``half-mirror" effect due to quantum tunneling \end{center}} 
& \parbox[c]{0.25\textwidth}{\begin{flushleft} electromagnetic energy of electromagnetic wave 
\end{flushleft}} 
&\parbox[c]{0.25\textwidth}{\begin{flushleft} \citet{press72,teukolsky74,lightman75} 
\end{flushleft}} \\ 
\enddata
\end{deluxetable}



\acknowledgments
We are grateful to Mika Koide for her helpful comments on this paper.

\end{document}